# Investigation of Precursor Superconducting State in $YBa_2Cu_3O_{7-\delta}$ through In-plane Optical Spectroscopy

Kegan Lee, Keisuke Kamiya, Masamichi Nakajima, Shigeki Miyasaka, Setsuko Tajima

Department of Physics, Osaka University, Toyonaka 560-0043, Japan

A precursor of superconductivity has been searched in the in-plane optical spectra of underdoped $YBa_2Cu_3O_y$, in which the previous c-axis optical spectra showed the presence of superconducting carriers at a temperature far above $T_c$.[1,2] By carefully subtracting the normal component from the imaginary part of conductivity $\sigma_2(\omega)$, we found a clear in-plane response of superconducting condensate at the temperature consistent with the c-axis optical data. This confirms that the precursory superconductivity developing with reducing a doping level is an intrinsic phenomenon in the cuprates.

The complete phase diagram for the high-temperature superconducting cuprates is still under intense debate. In particular, the origin of the pseudogap, as well as the existence of the precursor superconductivity, has been attracting much attention in recent years. While the pseudogap was previously thought to be synonymous with the precursor superconducting state,[3,4] many experiments indicated that the pseudogap is not a precursor of superconductivity but a competing order. For example, the c-axis optical studies[5] of Zn-doped $YBa_2Cu_3O_{6+\delta}$(YBCO) showed that the pseudogap temperature is insensitive to the Zn impurity. The gap opens even if the material is non-superconducting owing to impurity pair breaking[6] and thus appears to be independent of superconductivity. Zn-insensitive behavior of the pseudogap temperature has been reported by many research groups.[7-9]

Independent of the pseudogap, it has been revealed by the c-axis optical spectra of YBCO with various doping levels ($p$) that a superconducting condensate is present up to temperatures ($T_p$) way above $T_c$ (superconducting transition temperature) but lower than $T^*$ (pseudogap temperature).[1] This observation supports the c-axis ellipsometry results[2] which indicated the phonon softening at 320cm$^{-1}$ accompanied with the appearance of the transverse Josephson plasma mode at the temperature that coincides with $T_p$. Prior to the optical approaches, experiments probing the Nernst effect[10] and the torque magnetization[11] have also shown responses that fit the temperature scale of $T_p$ reasonably well. Although the transport experiments do not probe the superfluid density directly, the results provided compelling evidence of the presence of precursor superconductivity. Recently angle-resolved photoemission spectroscopy (ARPES) on $Bi_2Sr_2CaCu_2O_z$ has detected superconducting signals at the temperature higher than $T_c$ but lower than $T^*$.[12]

One of the problems is that so far the optical observations of precursory superconductivity have been limited to the c-axis direction, although a superconducting signal should be also observed in the ab-direction that is the main conduction path.[13] It is therefore imperative that we seek similar responses from the ab plane (in-plane) measurements to confirm that the c-axis results are genuinely due to the superconducting phenomenon. The problem of the in-plane optical measurement is that it is difficult to accurately estimate a superfluid density from the missing area in the real part of conductivity $\sigma_1(\omega)$ because with the lowering of the temperature, most of the spectral weight in the highly conductive in-plane spectrum is transferred to the lower frequency region that is out of the measurement range. In order to overcome this difficulty, we need to estimate a superconducting condensate from the imaginary part of conductivity, $\sigma_2(\omega)$. In this paper, we report the first observation of a superconducting condensate through a-axis optical spectroscopy that appears at a temperature which fits well with the $T_p$ determined from the c-axis optical measurements.

YBCO crystals used for this research were grown with a pulling technique (SRL-CP).[14] The as-grown crystals are then carefully cut into ~$3 \times 3 \times 1$ (or 0.5) mm$^3$ slabs. To control the oxygen contents, the two samples were annealed under oxygen gas flow for approximately two weeks at 675°C and 590°C, respectively. The surfaces of the samples were finely polished and detwinned under argon gas flow by applying a uniaxial pressure. Detwinning was confirmed by observing the surface through a polarized microscope and we estimate that more than 80% of the single crystal was successfully detwinned. The superconducting transition temperatures $T_c$ of the samples (63K and 77K respectively) were determined from magnetic susceptibility measurements with a magnetic field of 10 Oe. The superconductivity transitions were sharp with a transition width of about 3K, hence we defined the midpoint of the transition as the $T_c$s of the samples. Hereafter we call these two samples UD63K and UD77K, respectively. The doping levels $p$ ($p$=0.11 and 0.136) were then determined from the $T_c$ values, based on the $T_c$-$p$ relation published in the literature.[15] All of the present optical spectroscopy measurements were carried out in a He flow cryostat with light polarized in the a-direction using a Brukers Vertex 80V Fourier Transform Infrared spectrometer in the energy range from 80cm$^{-1}$ to 20000cm$^{-1}$. The spectrum above this range up to 350000cm$^{-1}$ was measured in the Ultraviolet Synchrotron Orbital Radiation (UVSOR) facility in Okazaki, Japan.

It is well established to use the Kramers-Kronig (KK) analysis on the reflectivity data to obtain the real and imaginary parts of optical conductivity, $\sigma_1(\omega, T)$ and $\sigma_2(\omega, T)$. To perform the KK analysis, data from 0cm$^{-1}$ to infinity is required but the range of measurement for our experiments is from 80 cm$^{-1}$ to approximately 317000 cm$^{-1}$. Hence for data above $T_c$, the low energy region is extrapolated using Hagen Rubens functions, while the $\alpha(1-\omega^2)$ approximation was used for extrapolating data below $T_c$. For the high energy extrapolation, the free carrier approximation $R \propto \omega^{-4}$ was used. The real optical conductivity $\sigma_1(\omega, T)$ can be represented as

$$\sigma_1(\omega, T) = \sigma_{1,n}(\omega, T) + \sigma_{1,s}(\omega, T), \quad (1a)$$
$$\sigma_{1,s}(\omega, T) = \delta(0)\omega_{ps}^2/8, \quad (1b)$$

where $\sigma_{1,n}$ and $\sigma_{1,s}$ are the non-superconducting and the superconducting component, respectively. The second equation indicates that all the superconducting spectral weights are condensed into a

Dirac delta peak at ω=0. Thus, assuming the Ferrell-Glover-Tinkham sum rule[16]

$$\int_0^\infty \sigma_1(\omega)d\omega = \omega_p^2/8, \qquad (2)$$

we are able to estimate the superfluid density $\omega_{ps}^2$ from the missing area of $\sigma_1(\omega)$,

$$\omega_{ps}^2(T) = 8\int_0^\infty [\sigma_{1,normal}(\omega) - \sigma_1(\omega,T)]\,d\omega. \quad (3)$$

Here, $\sigma_{1,normal}(\omega)$ is the conductivity in the normal state. For example, we can use the room temperature spectrum $\sigma_1(\omega)$ for $\sigma_{1,normal}(\omega)$. Since the high energy spectra are affected by the interband excitations, we need to set the cut-off frequency $\omega_{cut}$ for integral instead of ω=∞. The problem is that this approach is highly sensitive to the choice of the low energy extrapolation below the energy range of the measurements. This ω-region below 80cm$^{-1}$ shows very strong temperature dependence and possesses a significantly large spectral weight. As a result, the difference in two spectra with different extrapolations in eq.(3) would lead to a very large error and uncertainty. This may be one of the reasons why previous optical studies failed to detect a precursor signal of superconductivity[13].

Instead of this spectral weight analysis, in the present study, we employ a different approach which makes use of the imaginary part of conductivity $\sigma_2(\omega,T)$. $\sigma_2(\omega,T)$ can also be expressed in terms of its non-superconducting and superconducting components:

$$\sigma_2(\omega,T) = \sigma_{2,n}(\omega,T) + \sigma_{2,s}(\omega,T), \quad (4a)$$

$$\sigma_{2,s}(\omega,T) = \omega_{ps}^2/4\pi\omega. \qquad (4b)$$

This approach enables us to directly calculate the superconducting component by subtracting the non-superconducting component $\sigma_{2,n}(\omega,T)$ which can be obtained by performing KK transformation on the measured $\sigma_1(\omega,T)[=\sigma_{1,n}(\omega,T)\ at\ \omega>0]^{17)}$:

$$\sigma_{2,n}(\omega,T) = -\frac{2\omega}{\pi}P\int_0^\infty \frac{\sigma_{1,n}(\omega',T)}{\omega'^2-\omega^2}d\omega'. \quad (5)$$

Then, we are able to estimate the superfluid density using the following formula:

$$\omega_{ps}^2(T) = 4\pi\omega[\sigma_2(\omega,T) - \sigma_{2,n}(\omega,T)]. \quad (6)$$

Namely, the quantity on the right hand side should be ω-independent.

The two samples were cooled down to 4K and the optical reflectivity was measured at various temperatures between 4K and 293K. As clearly seen in Fig. 1, the reflectivity gradually increases as temperature is lowered. Below $T_c$, the low energy region of the reflectivity is close to unity. Several phonon peaks are clearly observable in the far-infrared region even though reflectivity is very high.

We then calculated $\omega[\sigma_2(\omega,T) - \sigma_{2,n}(\omega,T)]$ not only for the low temperatures below $T_c$ but also for the high temperatures above $T_c$, following the procedure described above. The results are

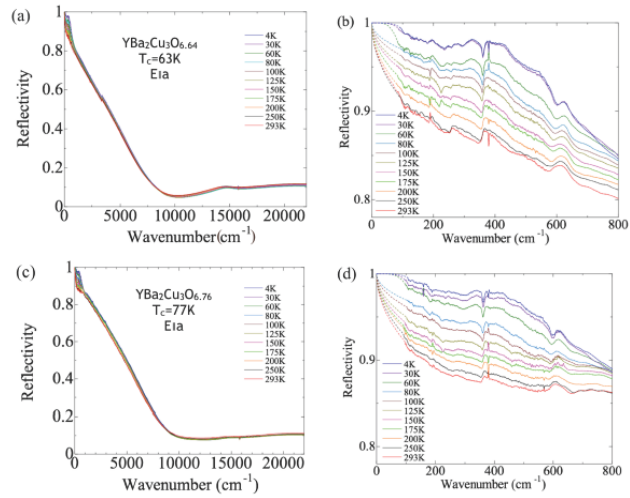

shown in

**Fig. 1** Reflectivity spectra of UD63K(a) and UD77K(c) over the full measured range. The expanded figures in the far-infrared region are also shown for UD63K(b) and UD77K(d).

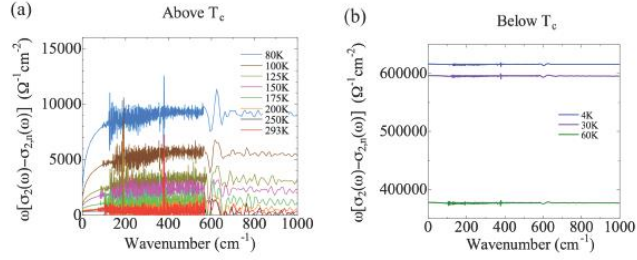

**Fig. 2** $\omega[\sigma_2(\omega,T) - \sigma_{2,n}(\omega,T)]$ of UD63K (a)above $T_c$ and (b)below $T_c$

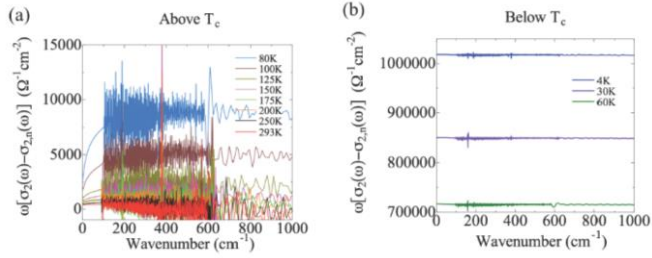

**Fig. 3** $\omega[\sigma_2(\omega,T) - \sigma_{2,n}(\omega,T)]$ of UD77K (a)above $T_c$ and (b)below $T_c$

Figs. 2 and 3. We note that although the fluctuation is relatively large, $\omega_{ps}^2$ is almost constant in the low energy region, indicating the relation of eq.(6) holds.

By calculating the average value of $\omega_{ps}^2(T)$ and estimating the magnitude of the error to be the width of the fluctuations, we plot $\omega_{ps}^2$ with error bars against $T$ in Fig. 4. Even with relatively large error bars, the superfluid density indisputably begins to have a finite value at temperatures much higher than $T_c$. We determine $T_p$ to be 170K and 130K for UD63K and UD77K, respectively. The value of $\omega_{ps}^2$ at the lowest temperature gives the penetration depth $\lambda_L$. For UD63K, it is estimated to be about 2400 Å, which agrees well with previous optical studies ($\lambda_L$=2800Å for a sample with doping level p=0.09[18]) and $\lambda_L$=1900Å for a sample with doping level p=0.10[13]).

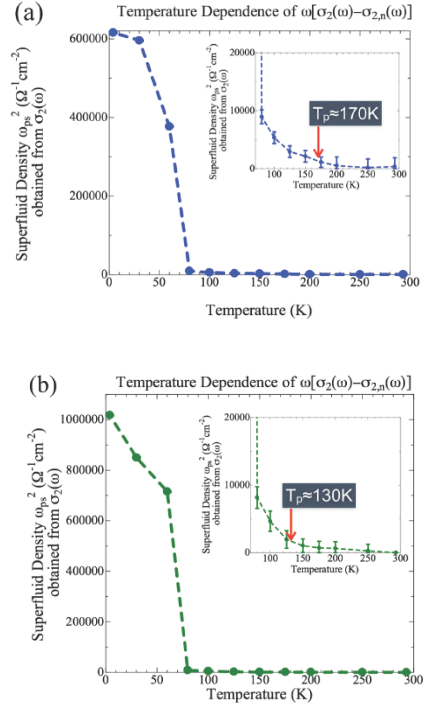

**Fig. 4** Temperature dependence of Superfluid Density $\omega_{ps}^2$ for (a)UD63K and (b)UD77K. The insets are the plots in the expanded scale.

From the comparison of $\omega_{ps}^2$ with the value at the lowest temperature, the superconducting volume fraction is estimated to be about 0.2- 0.3% near $T_p$, while ~2% just below $T_c$. This is smaller than the case of the c-axis spectra. Such a small volume fraction of precursory superconductivity may be the reason why this phenomenon has not been observed so far in the in-plane spectra. The recent ARPES study for $Bi_2Sr_2CaCu_2O_z$ demonstrated that the precursor superconductivity signal can be more easily detected when we go toward the anti-nodal direction of the Fermi surface.[19] This could be one of the origins of the difference in the precursor signal levels between the in-plane and c-axis measurements.

Many efforts have been devoted to find a superconducting signal, in particular, in microwave and THz regions[20-24]. All of these works successfully detected a superconducting fluctuation which is probably understood within the Ginsberg-Landau theory[25] because the onset temperature for this fluctuation is about 10-20K above $T_c$ that is well below $T_p$. We cannot specify the reason why they failed to detect a precursor of superconductivity far above $T_c$. One reason is that the signal in the in-plane direction is very small, as mentioned above. The other reason may be that both of $\sigma_1(\omega)$ and $\sigma_2(\omega)$ are dominated by the large normal components $\sigma_{1n}(\omega)$ and $\sigma_{2n}(\omega)$, which makes it difficult to extract a small precursor signal. In our analysis, the normal component is successfully subtracted by the Kramers-Kronig analysis.

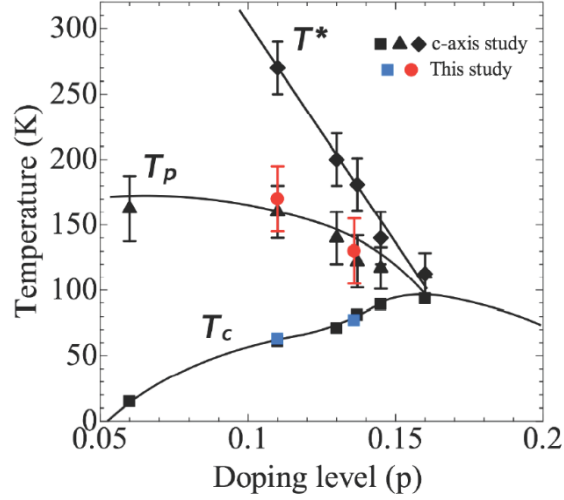

**Fig. 5** Phase diagram showing the doping dependence of $T_c$, $T^*$ and $T_p$ obtained from this study and c-axis studies[8].

Next we plot $T_p$ as a function of $p$ for the present two samples together with the results obtained from the c-axis optical studies[8] in Fig. 5. The error with regard to temperature is about $\pm 25$K, corresponding to the magnitude of our measurements' temperature intervals. The present a-axis results fit very well with the c-axis ones, providing compelling evidence for the existence of a precursor superconducting state in YBCO even though the volume fraction is small. We also recognize that $T_p$ is not very well-defined as the increase in superfluid density appears to be gradual. This suggests that with greater measurement precision, there is a possibility that we can observe precursor superconductivity at even higher temperatures. In other words, $T_p$ may be higher than what has been observed in our current experiments. The recently reported instant superconductivity at room temperature induced by femtosecond THz irradiation may be detecting the same precursory

superconductivity.[26]

The general trend is that $T_p$ increases as the doping level decreases. Although we did not specifically measure $T^*$ in our current experiments, many previous studies have verified that $T^*$ indeed increases with decreasing $p$, which is similar to the trend of $T_p$. Since it is established that the pseudogap is opening below $T^*$ including the $T$-region below $T_p$, we can conclude that a precursor of superconductivity coexists with the pseudogap. The problem is whether these two states are competing with each other or not. The opposite doping dependence of $T_c$ and $T^*$ suggests that the pseudogap is competing with superconductivity, whereas the doping dependence of $T_p$ is similar to that of $T^*$. If the pairing interaction originates from the Mott state, the increase of the superconducting transition temperature towards Mott insulator is reasonable. Then, one may consider that the original $T_c$ corresponding to $T_p$ increases with decreasing $p$ but is suppressed by the competing pseudogap, thus resulting in the actual bulk $T_c$. A similar $p$-dependence of $T_p$ and $T^*$ suggests that both are caused by the strong correlation that gives rise to a Mott insulator.

The plateau observed in the $T_c$ dome around 60K is often attributed to some sort of charge or spin ordering,[27] be it static or nematic, which is thought to suppress superconductivity. If this suppression phenomenon is true, it should be expected to suppress the precursor superconducting state in the same manner but this behavior is not observed in neither the present results nor in the c-axis optical studies. Although some reports also suggest that the charge-density wave (CDW) transition temperatures are very close to $T_p$ [27-29], the relation between CDW and precursor superconductivity is currently still not well understood.

In summary, a finite increase in superfluid density was observed in the in-plane optical spectra of two underdoped YBCO samples at the temperatures ($T_p$) much higher than $T_c$ but lower than $T^*$. These observations are consistent with the findings from the c-axis optical studies, although the precursor signal is weaker than the case of the c-axis spectra. Gradual growth of precursor signal suggests that the precursory superconductivity may persist up to much higher temperatures than $T_p$.

This work was supported by Grand-in-Aid for Scientific Research (Grant No. 24340083) from the MEXT, Japan.

---